\documentclass[aps,nofootinbib,superscriptaddress,twocolumn]{revtex4}
\usepackage{amstext,amsmath,amssymb,amsfonts,bbm}
\usepackage[latin1]{inputenc}
\usepackage{fancyhdr}
\usepackage{graphicx}
\usepackage{hyperref}
\usepackage{epstopdf}
\usepackage{amsthm}

\def\be{\begin{equation}}
\def\ee{\end{equation}}
\def\bes{\begin{eqnarray}}
\def\ees{\end{eqnarray}}

\newcommand{\bra}[1]{\ensuremath{\langle#1|}}
\newcommand{\ket}[1]{\ensuremath{|#1\rangle}}
\newcommand{\bk}[2]{{\langle#1\,|\,#2\rangle}}
\newcommand{\bek}[3]{{\langle#1\,|\,#2\,|\,#3\rangle}}

\theoremstyle{definition}
\theoremstyle{definition}

\begin{document}

\title{\large \bf Face amplitude of spinfoam quantum gravity}

\author{Eugenio Bianchi}     \affiliation{Centre de Physique Th\'eorique de Luminy\footnote{Unit\'e mixte de recherche (UMR 6207) du CNRS et des Universit\'es de Provence (Aix-Marseille I), de la M\'editerran\'ee (Aix-Marseille II) et du Sud (Toulon-Var); laboratoire affili\'e \`a la FRUMAM (FR 2291).}, Case 907, F-13288 Marseille, EU}

\author{Daniele Regoli}     \affiliation{Centre de Physique Th\'eorique de Luminy\footnote{Unit\'e mixte de recherche (UMR 6207) du CNRS et des Universit\'es de Provence (Aix-Marseille I), de la M\'editerran\'ee (Aix-Marseille II) et du Sud (Toulon-Var); laboratoire affili\'e \`a la FRUMAM (FR 2291).}, Case 907, F-13288 Marseille, EU}\affiliation{Dipartimento di Fisica Universit\`a di Bologna e INFN sezione di Bologna, via Irnerio 46, 40126 Bologna, Italy.}

\author{Carlo Rovelli}     \affiliation{Centre de Physique Th\'eorique de Luminy\footnote{Unit\'e mixte de recherche (UMR 6207) du CNRS et des Universit\'es de Provence (Aix-Marseille I), de la M\'editerran\'ee (Aix-Marseille II) et du Sud (Toulon-Var); laboratoire affili\'e \`a la FRUMAM (FR 2291).}, Case 907, F-13288 Marseille, EU}

\date{\small\today}

\begin{abstract}\noindent 
The structure of the boundary Hilbert-space and the condition that amplitudes behave appropriately under compositions determine the face amplitude of a spinfoam theory. In quantum gravity the face amplitude turns out to be simpler than originally thought. 
\end{abstract}

\maketitle

\section{Introduction}

A spinfoam sum over a given two-complex $\sigma$, formed by faces $f$ joining along edges $e$ in turn meeting at vertices $v$, is defined by the expression
\be
   Z_\sigma = \sum_{j_f, i_e}\ \prod_f d_{j_f}\ \prod_v A_v(j_f,i_e),
   \label{Z}
\ee
where $A_v(j_f,i_e)$ is the ``vertex amplitude" and $d_{j_f}$ is the ``face amplitude". The sum is over an assignment $j_f$ of an irreducible representation of a compact  group $G$ to each face $f$ and of an intertwiner $i_e$ to each edge $e$ of the two-complex. The expression \eqref{Z} is often viewed as a possible foundation for a background independent quantum theory of gravity \cite{Perez:2003vx}.  In particular, a vertex amplitude $A_v(j_f,i_e)$ that might define a quantum theory of gravity has been developed in \cite{Barrett:1997gw,Engle:2007uq,Engle:2007qf,Pereira:2007nh,Engle:2007wy,Livine:2007vk,Freidel:2007py, Kaminski:2009fm} and is today under intense investigation (see \cite{Rovelli:2010wq}).  But what about the ``measure factor" given by the face amplitude $d_{j_f}$?   What determines it? 

The uncertainty in determining the face amplitude has been repeatedly remarked \cite{Perez:2000fs,Perez:2000bf,Crane:2001as,Bojowald:2009im,Perini:2008pd,Pereira:2010}.  One way of fixing the face amplitude which can be found in the literature, for example, is to derive the sum \eqref{Z} for general relativity (GR) starting from the analogous sum for a topological $BF$ theory, and then implementing the constraints that reduce $BF$ to GR as suitable constraints on the states summed over.  For instance, in the Euclidean context GR is a constrained $SO(4)$ $BF$ theory. The state sum  \eqref{Z} is well understood for $SO(4)$ $BF$ theory: its face amplitude is the dimension of the $SO(4)$ representation  $(j_+,j_-)$. The simplicity constraint fixes this to be of the form $j_\pm= \gamma_\pm\; j_f$ where $ \gamma_\pm=\frac{1\pm\gamma}{2}$ and $\gamma$ is the Barbero-Immirzi parameter, and therefore 
\be
   d_{j_f} = (2j_++1)(2j_-+1)=(2\gamma_+j_f+1)(2\gamma_-j_f+1).
   \label{bffa}
\ee
However, doubts can be raised against this argument. For instance, Alexandrov \cite{Alexandrov:2010pg} has stressed the fact that the implementation of second class constraints into a Feynman path integral in general requires a modification of the measure, and here the face amplitude plays precisely the role of such measure, since $A_v\sim e^{i\,Action}$. Do we have an independent way of fixing the face amplitude?

Here we argue that the face amplitude is uniquely determined for any spinfoam sum of the form \eqref{Z} by three inputs: (a) the choice of the boundary Hilbert space, (b) the requirement that the composition law holds when gluing two-complexes; and (c) a particular ``locality" requirement, or, more precisely, a requirement on the local composition of group elements.  

We argue below that these requirements are implemented if $Z$ is given by the expression 
\be
Z_\sigma=\int dU_f^v\  \prod_v  A_v(U_f^v)\ \prod_f\ \delta(U_f^{v_1}...U_f^{v_k}), 
\label{main}
\ee
where $U_f^v\in G$,   $v_1...v_k$ are the vertices surrounding the face $f$, and $A_v(U_f^v)$ is the vertex amplitude $A_v(j_f,i_e)$ expressed in the group element basis \cite{Bianchi:2010}.  Then we show that this expression leads directly to \eqref{Z}, with arbitrary vertex amplitude, but a fixed choice of face amplitude, which turns out to be the dimension of the representation $j$ of the group $G$,  
\be
           d_j= {\rm dim}(j).
 \label{d}
\ee
In particular, for quantum gravity this implies that the $BF$ face amplitude \eqref{bffa} is ruled out, and should be replaced (both in the Euclidean and in the Lorentzian case) by the $SU(2)$ dimension
\be
           d_j= 2j+1.
\ee

Equation \eqref{main} is the key expression of this paper; we begin by showing that $SO(4)$ BF theory (the prototypical spinfoam model) can be expressed in this form (Section \ref{BFsec}).  Then we discuss the three requirements above and we show that   \eqref{main} implements these requirements. (Section \ref{inputs}).  
Finally we show that  \eqref{main} gives  \eqref{Z} with the face amplitude \eqref{d} (Section IV).

The problem of fixing the face amplitude has been discussed also by Bojowald and Perez in \cite{Bojowald:2009im}. Bojowald and Perez demand that the amplitude be invariant under suitable refinements of the two-complex. This request is strictly related to the composition law that we consider here, and the results we obtain are consistent with those of  \cite{Bojowald:2009im}.

\section{\emph{BF} theory}\label{BFsec}

\begin{figure}
\centering
\includegraphics[width=0.30\textwidth]{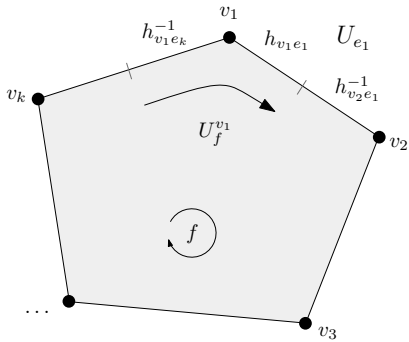}
\caption{Schematic definition of the group elements $h_{ve}$, $U^v_f$ and $U_e$ associated to a portion of a face $f$ of the two-complex.}
\label{fig1}
\end{figure}

It is well known  that the partition function \eqref{Z} for $BF$ theory can be rewritten in the form (see \cite{Perez:2003vx})
\be
Z_\sigma=\int dU_e\   \prod_f\ \delta(U_{e_1}...U_{e_k}), 
\label{ZBF}
\ee
where $U_e$ are group elements associated to the oriented \emph{edges} of $\sigma$, and $(e_1,...,e_k)$ are the edges that surround the face $f$.   Let us introduce group elements $h_{ve}$, labelled by a vertex $v$ and an adjacent edge $e$, such that 
\be
U_e =h_{ve}h_{v'e}^{-1}
\ee
where $v$ and $v'$ are the source and the target of the edge $e$  (see Figure \ref{fig1}). Then we can trivially rewrite \eqref{ZBF} as 
\be
Z_\sigma=\int dh_{ve} \prod_f\ \delta\!\left((h_{v_1e_1} h^{-1}_{v_2 e_1})\ ...\ (h_{v_k e_k} h^{-1}_{v_1 e_k})\right). 
\label{ZBF3}
\ee
Now define the group elements 
\be
U_f^v = h^{-1}_{ve} h_{ve'}
\ee
associated to a single vertex $v$ and two edges $e$ and $e'$ that emerge from $v$ and bound the face $f$ (see Figure \ref{fig1}). Using these, we can rewrite \eqref{ZBF} as 
\be
Z_\sigma=\int\! dh_{ve}  \int\! dU_f^v \,
\prod_{v,f^v} \delta(U_f^v,h^{-1}_{ve} h_{ve'})
\prod_f\ \delta(U_{f}^{v_1}...U_{f}^{v_k}),
\nonumber 
\ee
where the first product is over faces $f^v$ that belong to the vertx $v$, and then a product over all the vertices of the two-complex.

Notice that this expression has precisely the form \eqref{main}, where the vertex amplitude is
\be
A_v(U_f^v)=\int dh_{ve}  \prod_{f^v} \delta(U_f^v,h_{ve} h_{ve'}^{-1}),
\ee
which is the well-known expression of the 15j Wigner symbol (the vertex amplitude of $BF$ in the spin network basis) in the basis of the group elements. 

We have shown that the $BF$ theory spinfoam amplitude can be put in the form  \eqref{main}. We shall now argue that   \eqref{main} is the \emph{general} form of a local spinfoam model that obeys the composition law.

\section{Three inputs}\label{inputs}

(a) \emph{Hilbert space structure}. Equation \eqref{Z} is a coded expression to define the amplitudes
\be
   W_\sigma(j_l,i_n) = \sum_{j_f, i_e} \prod_f d_{j_f} \prod_v A_v(j_f,i_e;j_l,i_n), 
   \label{W}
\ee
defined for a two-complex $\sigma$ \emph{with boundary}, where the boundary graph $\Gamma=\partial\sigma$ if formed by links $l$ and nodes $n$. The spins $j_l$ are associated to the links $l$, as well as to the faces $f$ that are bounded by $l$; the intertwiners $i_n$ are associated to the nodes $n$, as well as to the edges $e$ that are bounded by $n$. The amplitude of the vertices that are adjacent to these boundary faces and edges depend also on the external variables $(j_l,i_n)$.  

In a quantum theory, the amplitude $W(j_l,i_n)$ must be interpreted as a (covariant) vector in a space $H_\Gamma$ of quantum states.\footnote{If $\Gamma$ has two disconnected components interpreted as ``in" and an ''out" spaces, then  $H_\Gamma$ can be identified as the tensor product of the ``in" and an ''out" spaces of non-relativistic quantum mechanics. In the general case,  $H_\Gamma$ is the boundary quantum state in the sense of the boundary formulation of quantum theory \cite{Rovelli,Oeckl:2003vu}.} We assume that this space has a Hilbert space structure, which we know.  In particular, we assume that 
\be
 {\cal H}_\Gamma= L_2[G^L,dU_l]
\label{n}
\ee
where $L$ is the number of links in $\Gamma$ and $dU_l$ is the Haar measure. Thus we can interpret \eqref{W} as
\be
   W_\sigma(j_l,i_n) = \bk{j_l,i_n}{W}
      \label{Wbraket}
\ee
where $\ket{j_l,i_n}$ is the spin network function 
\be
\bk{U_l}{j_l,i_n}=\psi_{j_l,i_n}(U_l)= \bigotimes_l  R^{j_l}(U_l)\cdot \bigotimes_n i_n.
\ee
Here $R^j(U_l)$ are the representation matrices in the representation $j$ and the $i$ form an orthonormal basis in the intertwiner space. See for instance \cite{Rovelli,Rovelli:2010wq} for details. Using the scalar product defined by \eqref{n}, we have 
\bes
\bk{j_l,i_n}{j'_l,i'_n}&=& \int  dU_l\ \overline{\psi_{j_l,i_n}(U_l)}\psi_{j'_l,i'_n}(U_l)
\nonumber\\
&=&
 \prod_l{\rm dim}(j_l)\; \delta_{j_lj'_l} \ \prod_n \delta_{i_ni'_n}.
\ees
where ${\rm dim}(j)$ is the dimension of the representation $j$. Therefore the spin-network functions $\psi_{j_l,i_n}(U_l)$ are not normalized.  (These ${\rm dim}(j)$ normalization factors are due to the convention chosen: they have nothing to do with the dimension of the representation that appears in \eqref{d}.) The resolution of the identity in this basis is 
\be
1\!\!1=\sum_{j_l,i_n}\ \big(\, \text{\footnotesize $\displaystyle \prod_l$}\, {\rm dim}(j_l)\big)\ \ket{j_l,i_n}\bra{j_l,i_n}.
\ee\\

(b) \emph{Composition law.} In non relativistic quantum mechanics, if $U(t_1,t_0)$ is the evolution operator from time $t_0$ to time $t_1$, the composition law reads
\be
U(t_2,t_0)=U(t_2,t_1)U(t_1,t_0). 
\ee
That is, if $\ket{n}$ is an orthonormal basis,
\be
\bek{f}{\!U(t_2,t_0)\!}{i}=
\sum_{n}\bek{f}{\!U(t_2,t_1)\!}{n}\bek{n}{\!U(t_1,t_0)\!}{i}.\nonumber
\ee
Let us write an analogous condition of the spinfoam sum. Consider for simplicity a 
two-complex $\sigma=\sigma_1\cup\sigma_2$ without boundary, obtained by gluing two two-complexes $\sigma_1$ and $\sigma_2$ along their common boundary $\Gamma$. Then we require that $W$ satisfies the composition law 
\be
Z_{\sigma_1\cup\sigma_2}=\bk{W_{\sigma_2}}{W_{\sigma_1}},
\ee
as discussed by Atiyah in \cite{Atiyah:1989vu}. Notice that to formulate this condition we need the Hilbert space structure in the space of the boundary states. \\

\begin{figure}
\centering
\includegraphics[width=0.30\textwidth]{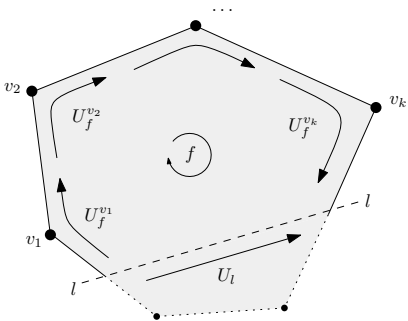}
\caption{Cutting of a face of the two-complex. The holonomy $U_l$ is attached to a link of the boundary spin-network and satisfies equation \eqref{locality}.}
\label{fig2}
\end{figure}

(c) \emph{Locality.} As a vector in $H_\Gamma$, the amplitude $W(j_l,i_n)$ can be expressed on the group element basis
\bes
W(U_l)&=&\bk{U_l}{W}\\
&=&\sum_{j_l,i_n} \; \big(\,\text{\footnotesize $\displaystyle \prod_l$}\, {\rm dim}(j_l)\big)\ \psi_{j_l,i_n}(U_l) W(j_l,i_n). \nonumber
\ees
Similarly, the vertex amplitude can be expanded in the group element basis
\bes
A_v(U_f^v)&=&\bk{U_f^v}{A_v} \label{va}\\
&=&\sum_{j_f^v,i_n^v} \; \big(\,\text{\footnotesize $\displaystyle \prod_{f^v}$}\, {\rm dim}(j_f^v)\big)\ \psi_{j_f^v,i_n^v}(U_f^v) A_v(j_f^v,i_n^v).\nonumber
\ees
Notice that here the group element $U_f^v$ and the spin $j_f^v$ are associated to a vertex $v$ and a face $f$ adjacent to $v$. Similarly, the intertwiner $i_n^v$ is associated to a vertex $v$ and a node $n$ adjacent to $v$. Consider a boundary link $l$ that bounds a face $f$ (see Figure \ref{fig2}). Let $v_1...v_k$ be the vertices that are adjacent to this face.  We say that the model is local if the relation between the boundary group element $U_l$ and the vertices group elements $U_f^v$ is given by 
\be
U_l=U_f^{v_1}\,...\ U_f^{v_k}.
\label{locality}
\ee
In other words: if the boundary group element is simply the product of the group elements around the face. 

\vspace{1em}

Notice that a spinfoam model defined by \eqref{main} is local and satisfies composition law in the sense above.  In fact, \eqref{main} generalizes immediately to 
\bes
W_\sigma(U_l)&=&\int dU_f^v\  \prod_v  A_v(U_f^v)\ \prod_{{\rm internal}\ f}\ \delta(U_f^{v_1}...U_f^{v_k})\nonumber\\
&& \ \ \times\ \ \prod_{{\rm external}\ f}\ \delta(U_f^{v_1}...U_f^{v_k}U^{-1}_l).
\label{main2}
\ees
Here the first product over $f$ is over the (``internal") faces that do not have an external boundary; while the second is over the (``external") faces $f$ that are also bounded by
the vertices $v_1, ..., v_k$ and by the the link $l$.  It is immediate to see that locality is implemented, since the second delta enforces the locality condition \eqref{locality}.

Furthermore, when gluing two amplitudes along a common boundary we have immediately that 
\be
\int dU_l\ \overline{W_{\sigma_1}(U_l)}\ W_{\sigma_2}(U_l)=Z_{\sigma_1\cup\sigma_2}
\ee
because the two delta functions containing $U_l$ collapse into a single delta function associated to the face $l$, which becomes internal.  

Thus, \eqref{main} is a general form of the amplitude where these conditions hold.

In \cite{Bojowald:2009im}, Bojowald and Perez have considered the possibility of fixing the face amplitude by requiring the amplitude of a given spin/intertwiner configuration to be equal to the amplitude of the same spin/intertwiner configuration on a finer two-simplex where additional faces carry the trivial representation.  This requirement imply essentially that the amplitude does not change by splitting a face into two faces. It is easy to see that \eqref{main} satisfies this condition. Therefore \eqref{main} satisfies also the Bojowald-Perez condition.

\section{Face amplitude}

Finally, let us show that \eqref{main} implies \eqref{Z} \emph{and} \eqref{d}. To this purpose, it is sufficient to insert \eqref{va} into  \eqref{main}. This gives
\bes
Z_\sigma&=&\!\int dU_f^v\  \prod_v  \sum_{j_f^v,i_n^v} \big(\,\text{\footnotesize $\displaystyle \prod_{f^v}$}\, {\rm dim}(j_f^v)\big)\ \psi_{j_f^v,i_n^v}(U_f^v) A_v(j_f^v,i_n^v)\nonumber\\
&&\ \ \times \   \prod_f\ \delta(U_f^{v_1}...U_f^{v_k}).
\label{main22}
\ees
Expand then the delta function in a sum over characters 
\bes
Z_\sigma&=&\!\int dU_f^v\  \prod_v  \sum_{j_f^v,i_n^v}\! \big(\,\text{\footnotesize $\displaystyle \prod_{f^v}$}\, {\rm dim}(j_f^v)\big)\ \psi_{j_f^v,i_n^v}(U_f^v) A_v(j_f^v,i_n^v)\nonumber\\
&&\ \ \times \   \prod_f\ \sum_{j_f}{\rm dim}(j_f)\ {\rm Tr}(R^{j_f}(U_f^{v_1})\cdots R^{j_f}(U_f^{v_k})).
\nonumber
\ees
We can now perform the group integrals.  Each $U_f^v$ appears precisely twice in the integral: once in the sum over $j_f^v$ and the other in the sum over $j_f$. Each integration gives a delta function $\delta_{j_f^v,j_f}$, which can be used to kill the sum over $j_f^v$ dropping the $v$ subscript. Following the contraction path of the indices, it is easy to see that these contract the two intertwiners at the opposite side of each edge. Since intertwiners are orthonormal, this gives a delta function $\delta_{i_n^v,i_n^{v'}}$ which reduces the sums over intertwiners to a single sum over $i_n:=i_n^v=i_n^{v'}$. Bringing everything together, and noticing that the ${\rm dim}(j)$ factor from the group integrations cancels the one in the integral, we have 
\be
Z_\sigma=\sum_{j_f\, i_n}\ \prod_f {\rm dim}(j_f)\ \prod_v  A_v(j_f^v,i_n^v).
\ee
This is precisely equation \eqref{Z}, with the face amplitude given by \eqref{d}.

Notice that the face amplitude is well defined, in the sense that it cannot be absorbed into the vertex amplitude (as any \emph{edge} amplitude can). The reason is that any factor in the vertex amplitude depending on the spin of the face contributes to the total amplitude at a power $k$, where $k$ is the number of sides of the face. The face amplitude, instead, is a contribution to the total amplitude that does not depend on $k$. This is also the reason why the normalization chosen for the spinfoam basis does not affect the present discussion: it affects the expression for the vertex amplitude, not that for the face amplitude.

By an analogous calculation one can show that the same result holds for the amplitudes $W$: equation \eqref{W} follows from \eqref{main2} expanded on a spin network basis. 

\vspace{3em}

In conclusion, we have shown that the general form \eqref{main} of the partition function, which implements locality and the composition law, implies that the face amplitude of the spinfoam model is given by the dimension of the representation of the group $G$ which appears in the boundary scalar product (\ref{n}). 

In general relativity, in both the Euclidean and the Lorentzian cases, the boundary space is 
\bes
 {\cal H}_\Gamma &=& L_2[SU(2)^L, {\rm d}U_l],
\ees
therefore the face amplitude is $d_j={\rm dim}_{SU(2)}(j)=2j+1$, and not the $SO(4)$ dimension \eqref{bffa}, as previously supposed.  

Notice that such $d_j=2j+1$ amplitude defines a theory that is far less divergent than the theory defined by \eqref{bffa}. In fact, the potential divergence of a bubble is suppressed by a power of $j$ with respect to \eqref{bffa}.  In \cite{Perini:2008pd}, it has been shown that  the $d_j=2j+1$ face amplitude yields a \emph{finite} main radiative correction to a five-valent vertex if all external legs set to zero.


\section*{Acknowledgements}
We thank Elena Magliaro, Antonino Marcian\`o, Claudio Perini, and Simone Speziale for useful discussions. EB gratefully acknowledges funding for this work through a Marie Curie Fellowship of the European Union.

\providecommand{\href}[2]{#2}\begingroup\raggedright\endgroup

\end{document}